\documentclass{article}
\usepackage{preprint,times}

\usepackage{multirow}
\usepackage{wrapfig}

\usepackage[utf8]{inputenc} % allow utf-8 input
\usepackage[T1]{fontenc}    % use 8-bit T1 fonts
\usepackage{hyperref}       % hyperlinks
\usepackage{url}            % simple URL typesetting
\usepackage{booktabs}       % professional-quality tables
\usepackage{amsfonts}       % blackboard math symbols
\usepackage{nicefrac}       % compact symbols for 1/2, etc.
\usepackage{microtype}      % microtypography
\usepackage{xcolor}         % colors
\usepackage{graphicx}       % figures
\usepackage{amsmath}        % math environments
\usepackage{subcaption}     % subfigures
\usepackage{arydshln}
\usepackage{colortbl}

\iclrfinalcopy
\title{Multi-agent Collaboration with State Management}

\author{Mengyang Liu$^{1,2}$, Taozhi Chen$^{3}$, Zhenhua Xu$^{4}$, Xue Jiang$^{4}$, Yihong Dong$^{4}$\\
$^1$ Shanghai Jiaotong University \quad $^2$ Cortices AI\nocite{CorticesAI} \quad $^3$ Emory University \\
$^4$ Peking University\\
\{mengyangliu912, chentaozhi313, EthanDongyh\}@gmail.com
}

\begin{document}

\maketitle

\begin{abstract}

Recent advances in multi-agent systems have shown great potential for solving complex tasks. However, when multiple agents edit a shared codebase concurrently, their changes can silently conflict and inconsistent views lead to integration failures. Existing multi-agent systems address this through workspace isolation (e.g., one git worktree per agent), but this defers conflict resolution to a post-hoc merge step where recovery is expensive. In this paper, we propose STORM, i.e., STate-ORiented Management for multi-agent collaboration. Specifically, STORM manages agent states by mediating their interactions with the shared workspace, ensuring that each agent operates on a consistent view of the codebase and that conflicting edits are detected and resolved at write time. We evaluate STORM on Commit0 and PaperBench across multiple LLMs. STORM outperforms the git-worktree-based multi-agent baseline by +18.7 on Commit0-Lite and +1.4 on PaperBench, while achieving comparable or better cost efficiency. Combined with single-agent runs, STORM reaches highest scores of 87.6 and 78.2 on the two benchmarks respectively, suggesting that explicit state management is a more effective foundation for multi-agent collaboration than workspace isolation. STORM can also be plugged into any multi-agent system seamlessly.\footnote{Our code is available at \url{https://github.com/dreamyang-liu/STORM}}.

\end{abstract}

\section{Introduction}
\label{sec:intro}

Multiple LLM agents working in parallel can solve tasks that are too large for any single agent to finish within its iteration budget~\citep{selfcollaboration, chatdev,metagpt,caid,coral}. In software engineering, agents can implement different modules concurrently; in scientific research, they parallelize experimental setups. But running agents in parallel on a shared workspace raises a question: when one agent edits a file, how do we know that its assumptions about the rest of the codebase are still valid?

We treat this as a state management problem. An agent interacts with its workspace through file reads and writes. When it modifies a file, its reasoning depends not just on that file but on context from other files it has read (dependencies, interfaces, specifications). The edit is only safe when those context files have not changed since the agent last read them. This is a local consistency requirement: the agent does not need a frozen snapshot of the entire workspace, just assurance that the specific files informing its current edit are up to date. 

Existing multi-agent systems mostly avoid this problem by giving each agent its own workspace (e.g., a git worktree) and merging afterward~\citep{caid,coral,statsclaw}. Isolation prevents interference during editing but pushes all conflict resolution to the merge step, after agents have already committed to potentially incompatible designs. Textual merge conflicts are easy to spot; semantic conflicts, where both sides compile individually but break when combined, are harder, and current tooling cannot resolve them automatically.

In this paper, we propose STORM (STate-ORiented Management), a state management framework for multi-agent collaboration. STORM mediates each agent's file reads and writes. Before accepting a write, it checks whether the agent's view of the target file and its context dependencies is still current. If another agent has modified any of those files in the interim, the write is rejected and the agent receives the updated content so it can retry from a correct baseline.

Our main contributions can be attributed as:
\begin{enumerate}
    \item A formulation of multi-agent state management as file-level context consistency: an agent's write is valid only if the target file and its read dependencies have not been modified since the agent last observed them.
    \item STORM, an architecture-agnostic state management framework for multi-agent collaboration that enforces local state consistency at write time, detecting conflicts immediately and enabling agents to retry from a correct baseline without workspace isolation.
    \item Empirical validation on Commit0-Lite\citep{commit0} and PaperBench\citep{PaperBench} with Sonnet~4.6. On Commit0-Lite, STORM achieves 82.5\% macro pass rate and 46.2\% weighted pass rate, outperforming single-agent (66.4\% / 20.7\%) and GitWorktree (63.8\% / 24.6\%) and we observe similar gain on Deepseek and Qwen model. On PaperBench, STORM scores 74.1 vs.\ 72.7 (GitWorktree) and 68.7 (single-agent).
\end{enumerate}

\section{STORM}
\label{sec:method}

Mainstream multi-agent systems give each agent its own git worktree~\citep{caid,coral,statsclaw} so that agents cannot interfere with each other while working. The cost is paid later: once agents finish their individual task, their branches must be merged back together. If two agents edit the same file, or made incompatible design choices about a shared interface, the merge fails. Worse, because each agent worked in isolation without seeing what others were doing, these conflicts tend to compound. Agent~A writes a helper assuming a certain signature; Agent~B changes that signature in its own branch; code that depends on both is now broken in a way that neither branch exhibits alone.

We avoid this by putting all agents in the same workspace. The key insight is that an agent does not need a globally consistent view of the entire repository to produce a correct edit. It only needs the files it has actually read to remain unchanged while it reasons. We call this property \emph{local state consistency} and formalize it in Section~\ref{sec:method-formulation}. In practice, even with disjoint task assignments, agents sometimes need to edit the same file (e.g., two agents each implementing different functions in a shared module). Most of their edits do not interact, but at the boundaries where they do, agents need a way to exchange information. STORM addresses this with two mechanisms: \emph{write-time conflict control} (Section~\ref{sec:method-occ}), which rejects a write whenever the agent's local view has gone stale and lets it retry with fresh context, and \emph{intent annotations} (Section~\ref{sec:method-intent}), structured comments that agents leave in the code so that when another agent reads the same file, it can see not just the raw code but the intent behind it, enabling coordination at these shared boundaries without explicit messaging.

\subsection{Local State Consistency}
\label{sec:method-formulation}

An LLM agent does not need a frozen snapshot of the entire workspace to produce a correct edit. It only needs the files it has actually read to remain unchanged while it reasons. We formalize this as \emph{local state consistency}.

\paragraph{Workspace and agents.} Let the workspace be a set of versioned files $\mathcal{W} = \{(f, v_f) \mid f \in \mathcal{F}\}$, where $v_f \in \mathbb{N}$ is the current version of file $f$. A manager agent $M$ decomposes a task $T$ into sub-tasks $\{\tau_1, \ldots, \tau_k\}$ and assigns each to an engineer agent $a_i$ together with a \emph{primary file set} $F_i \subseteq \mathcal{F}$:
\begin{equation}
  M: T \;\longrightarrow\; \{(\tau_i, F_i, a_i)\}_{i=1}^{k}, \quad \text{where } F_i \cap F_j = \emptyset \;\text{ for } i \neq j.
\end{equation}
The disjoint assignment reduces but does not eliminate conflicts: agents may still read or edit files outside their primary set (e.g., a shared utility or a common import).

\paragraph{Agent local state.} As agent $a_i$ works on $\tau_i$, it accumulates a \emph{read snapshot} $S_i$ recording every file it has observed and the version at observation time:
\begin{equation}
  S_i = \{(g, v_g^{\text{obs}}) \mid a_i \text{ has read } g\}.
\end{equation}
When $a_i$ issues a write to file $f$ producing new content $c'$, its generation depends only on $S_i$, the local context the LLM has seen, not on the full workspace state. This is the key asymmetry we exploit: correctness requires consistency of $S_i$, not of $\mathcal{W}$.

In practice, each task $\tau_i$ requires accessing (reading or modifying) a set of files $A_i \subseteq \mathcal{F}$ that may extend beyond the assigned $F_i$. When two agents' access sets overlap ($A_i \cap A_j \neq \emptyset$), the shared files form a boundary where conflicts may arise. STORM only needs to coordinate at these boundaries, leaving the non-overlapping majority of work fully parallel.

\paragraph{Write validity.} A write $(a_i, f, c')$ is \emph{valid} if and only if the agent's local state is still consistent with the current workspace:
\begin{equation}
\label{eq:validity}
  \forall\, (g, v_g^{\text{obs}}) \in S_i: \quad v_g^{\text{obs}} = v_g^{\text{cur}}.
\end{equation}
That is, no file that $a_i$ has read has been modified since its observation. A valid write is applied atomically: $v_f \leftarrow v_f + 1$ and the content of $f$ is updated to $c'$.

\paragraph{Conflict.} A write is \emph{conflicting} when Eq.~\ref{eq:validity} is violated. Two cases arise:
\begin{itemize}
  \item \textbf{Direct conflict}: the target file itself was updated ($v_f^{\text{obs}} < v_f^{\text{cur}}$), meaning another agent wrote to $f$ concurrently.
  \item \textbf{Stale dependency}: a dependency file $g \neq f$ was updated ($v_g^{\text{obs}} < v_g^{\text{cur}}$), meaning $a_i$'s reasoning may rest on outdated context.
\end{itemize}
In both cases, STORM rejects the write and returns the current state, enabling $a_i$ to refresh $S_i$ and retry from a correct baseline. Section~\ref{sec:method-occ} details the mechanism.

\begin{figure}[t]
  \centering
  \includegraphics[width=\linewidth]{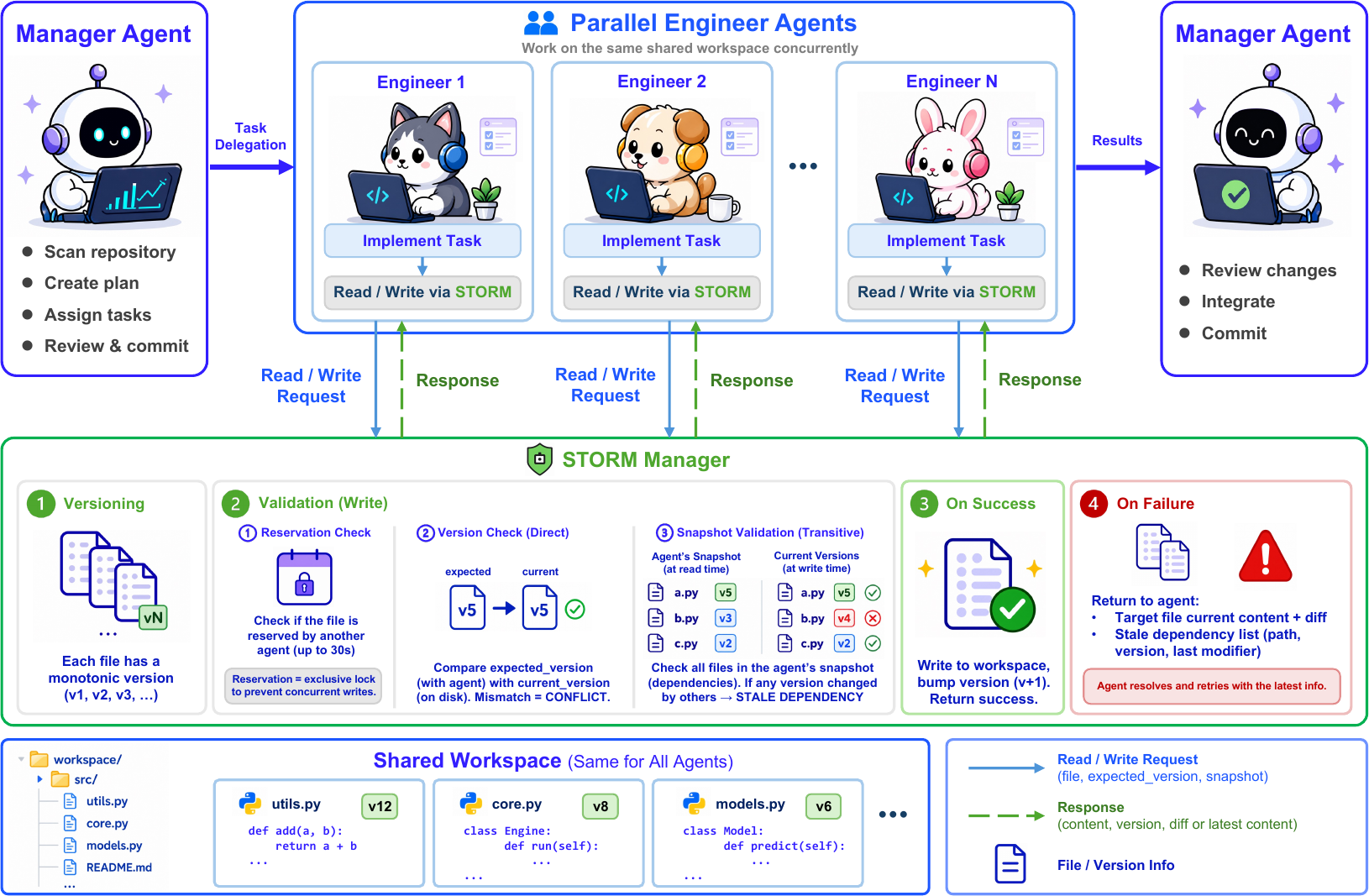}
  \caption{System architecture. The manager analyzes the repository, delegates tasks to parallel engineers, and commits their work. All engineers share one workspace; the STORM manager mediates file operations to detect and resolve conflicts.}
  \label{fig:architecture}
\end{figure}

Figure~\ref{fig:architecture} shows the architecture. A single manager agent reads the repository, partitions work into sub-tasks scoped to disjoint file sets, assigns each to an engineer, reviews diffs after engineers finish, runs tests, and commits accepted changes. Only the manager commits. Each engineer receives a scoped task (e.g., ``implement all functions in \texttt{tensor\_ops.py}'') and accesses the workspace only through the STORM-mediated \texttt{file\_editor}. Engineers do not communicate directly; coordination happens through the shared codebase and intent annotations (Section~\ref{sec:method-intent}).

\subsection{Write-time Conflict Control}
\label{sec:method-occ}

STORM enforces the validity condition in Eq.~\ref{eq:validity} via a mechanism inspired by optimistic concurrency control~\citep{kung1981occ}. The key observation is that in a well-decomposed task, most concurrent edits touch different files; the system lets all operations proceed without blocking and only intervenes when a conflict actually occurs.

\paragraph{Implementation.} Each file maintains a monotonically increasing version counter ($v_f$, starting at 1). Every \texttt{file\_editor} read returns the content together with $v_f$; every write must declare the expected version. The STORM layer validates the write against Eq.~\ref{eq:validity} by comparing the agent's full read snapshot $S_i$ to the current workspace state. If validation passes, the write is applied atomically ($v_f \leftarrow v_f + 1$). If it fails, the write is rejected.

\paragraph{Rejection payload.} On rejection, STORM returns: (1) the current content of the target file, (2) a unified diff showing what changed since the agent's last read (for direct conflicts), and (3) a list of stale dependencies with their version deltas. This gives the LLM enough context to re-plan from the current baseline without needing to re-read every file.

\paragraph{Reservation.} After a rejection, a short reservation is granted to the rejected agent on the target file. This prevents repeated alternating conflicts where two agents each invalidate the other in a tight loop.

\subsection{Intent annotations}
\label{sec:method-intent}

Version tracking catches file-level conflicts but not semantic ones. Two agents might implement the same helper with different signatures, or make incompatible assumptions about a shared data structure. To reduce this, engineers annotate their code with structured intent comments:

\begin{verbatim}
  # engineer_1: validate numeric inputs before summing
  def add(a, b):
      if not isinstance(a, (int, float)):
          raise TypeError("a must be numeric")
      return a + b
\end{verbatim}

Each comment identifies the author and describes what the block accomplishes. Engineers preserve annotations left by other agents unless their task requires changing the annotated block. When another agent reads the file, these comments provide a lightweight channel for semantic coordination: agents can see what others have done and avoid duplicate or conflicting work. The convention is injected into each engineer's system prompt automatically.

\section{Experiments}
\label{sec:exp}

\begin{table}[t]
  \caption{Aggregated results across Commit0-Lite and PaperBench Code-Dev. Best per model in \textbf{bold}, second best \underline{underlined}.}
  \label{tab:main-results}
  \centering
  \resizebox{\textwidth}{!}{
  \begin{tabular}{@{}l rrrr rrr@{\hspace{12pt}}@{}}
    \toprule
    & \multicolumn{4}{c}{\textbf{Commit0-Lite}} & \multicolumn{3}{c}{\textbf{PaperBench}} \\
    \cmidrule(lr){2-5} \cmidrule(lr){6-8}
    & Score\textsubscript{w} $\uparrow$ & Score $\uparrow$ & Cost\textsubscript{eff} $\downarrow$ & Time\textsubscript{eff} $\downarrow$ & Score $\uparrow$ & Cost\textsubscript{eff} $\downarrow$ & Time\textsubscript{eff} $\downarrow$ \\
    \midrule
    \multicolumn{8}{@{}l}{\cellcolor{gray!12}\textbf{Claude Sonnet 4.6}} \\
    \quad Single-Agent           & 20.7 & 66.4 & \textbf{3.2} & \underline{18.8} & 68.7 & \textbf{12.5} & \textbf{11.6} \\
    \quad GitWorktree            & 24.6 & 63.8 & 8.6 & 20.9 & 72.7 & 17.2 & \underline{22.5} \\
    \quad GitWorktree-Combined   & 31.4 & 78.6 & 8.1 & 25.7 & \underline{76.6} & 22.7 & 30.8 \\
    \quad STORM                  & \underline{46.2} & \underline{82.5} & \underline{6.3} & \textbf{16.8} & 74.1 & \underline{12.6} & 24.1 \\
    \quad STORM-Combined         & \textbf{49.2} & \textbf{87.6} & \underline{6.3} & 21.1 & \textbf{78.2} & 17.3 & 31.4 \\
    \multicolumn{8}{@{}l}{\cellcolor{gray!12}\textbf{Qwen 3.6 Plus}} \\
    \quad Single-Agent           & 34.0 & 75.3 & \textbf{1.3} & \underline{13.0} & 47.7 & \textbf{4.9} & \textbf{16.1} \\
    \quad GitWorktree            & 16.7 & 57.4 & 6.9 & 56.0 & 51.6 & 13.4 & 36.0 \\
    \quad GitWorktree-Combined   & 36.3 & \underline{83.0} & 3.7 & 31.1 & 55.4 & 12.1 & 43.1 \\
    \quad STORM                  & \underline{61.4} & 70.5 & 2.5 & \textbf{12.5} & \underline{55.0} & \underline{8.2} & \underline{21.0} \\
    \quad STORM-Combined         & \textbf{76.2} & \textbf{88.2} & \underline{2.1} & 13.2 & \textbf{57.0} & 10.7 & 29.6 \\
    \multicolumn{8}{@{}l}{\cellcolor{gray!12}\textbf{DeepSeek V4 Pro}} \\
    \quad Single-Agent           & 26.8 & 65.2 & \textbf{1.8} & \textbf{31.3} & 62.9 & \textbf{4.1} & \textbf{16.5} \\
    \quad GitWorktree            & 18.5 & 44.0 & 3.8 & 45.1 & 55.8 & 8.8 & 35.7 \\
    \quad GitWorktree-Combined   & 30.9 & \underline{75.2} & 3.8 & 53.9 & 60.8 & 9.1 & 41.3 \\
    \quad STORM                  & \underline{32.3} & 63.2 & \underline{3.0} & \underline{39.0} & \underline{66.5} & \underline{9.7} & \underline{35.5} \\
    \quad STORM-Combined         & \textbf{41.3} & \textbf{77.7} & 3.4 & 49.0 & \textbf{68.3} & 10.9 & 41.7 \\
    \bottomrule
  \end{tabular}}
\end{table}

We evaluate on two agent benchmarks: \textbf{Commit0-Lite}\citep{commit0}, and \textbf{PaperBench}\citep{PaperBench}. We compare five configurations across three LLMs (Claude Sonnet 4.6, Qwen 3.6 Plus, DeepSeek V4 Pro): (1)\textbf{Single-agent} with 100 iterations; (2)\textbf{GitWorktree}\citep{caid}, engineers in isolated worktrees merged after completion; (3)\textbf{STORM}, a manager with engineers sharing one workspace; and Combined variants that take the per-task best of single-agent and multi-agent runs. Commit0-Lite uses 4 engineers; PaperBench uses 2 engineers with 2 rounds of delegation each.

\textbf{Evaluation.} For Commit0-Lite, we run \texttt{pytest} on the final workspace and report \textbf{Score\textsubscript{w}} (total tests passed / total tests) and \textbf{Score} (mean per-repository pass rate). For PaperBench, we use the Code-Dev subset following~\citet{caid} due to cost constraints: the LLM judge (Sonnet~4.6) grades only ``Code Development'' nodes in the rubric tree, evaluating whether the submitted source code correctly implements each criterion without requiring experiment execution. We report \textbf{Score} (mean per-paper judge score $\times 100$). For both benchmarks, we report \textbf{Cost\textsubscript{eff}} (total cost / Score, lower is better) and \textbf{Time\textsubscript{eff}} (total wall-clock minutes across all tasks / Score, lower is better).

\textbf{Implementation.} All agents run on OpenHands~\citep{openhands}. Multi-agent runs follow a two-round protocol: in the first round the manager decomposes the task and dispatches engineers in parallel (each with 80 iterations); in the second round it reviews outputs, runs tests, and may reassign failed sub-tasks for one retry. The manager gets 50 iterations total. Detailed setup is in Appendix~\ref{appendix:setup}.

\subsection{Main results}

Table~\ref{tab:main-results} reports results on both Commit0-Lite (4 engineers) and PaperBench Code-Dev (2 engineers) across all three models. STORM achieves the \textbf{highest weighted score on Commit0-Lite across all models} (\textbf{46.2}, \textbf{61.4}, \textbf{32.3} for Sonnet, Qwen, DeepSeek), with gains concentrated on large repositories with cross-file dependencies where shared-workspace coordination matters most. GitWorktree performs worst across the board, dropping up to 18 points below single-agent on Commit0-Lite. On PaperBench, STORM consistently outperforms both single-agent and GitWorktree for all three models: \textbf{74.1} vs.\ 68.7 single-agent (Sonnet), \textbf{55.0} vs.\ 47.7 (Qwen), and \textbf{66.5} vs.\ 62.9 (DeepSeek), demonstrating that the manager's task decomposition and priority-driven delegation effectively covers more rubric criteria than a single agent working alone.

The single agent wins on cost-efficiency across all models due to zero coordination overhead, but STORM closes this gap while achieving substantially higher scores. STORM-Combined (per-paper best of single-agent and STORM) is the top-scoring configuration for every model on both benchmarks, reaching \textbf{78.2} on PaperBench (Sonnet), \textbf{57.0} (Qwen), and \textbf{68.3} (DeepSeek), confirming that the two approaches are \textbf{complementary}: single-agent excels on papers where one focused agent can cover most criteria, while STORM's parallel delegation captures broader coverage on complex multi-component papers. STORM also achieves better time-efficiency than GitWorktree on all models (e.g., \textbf{16.8} vs.\ 20.9 on Sonnet Commit0; \textbf{21.0} vs.\ 36.0 on Qwen PaperBench) because conflicts are resolved incrementally rather than in a costly merge step. Per-paper results are in Appendix~\ref{app:full-results}.

\begin{figure}[t]
  \centering
  \includegraphics[width=\linewidth]{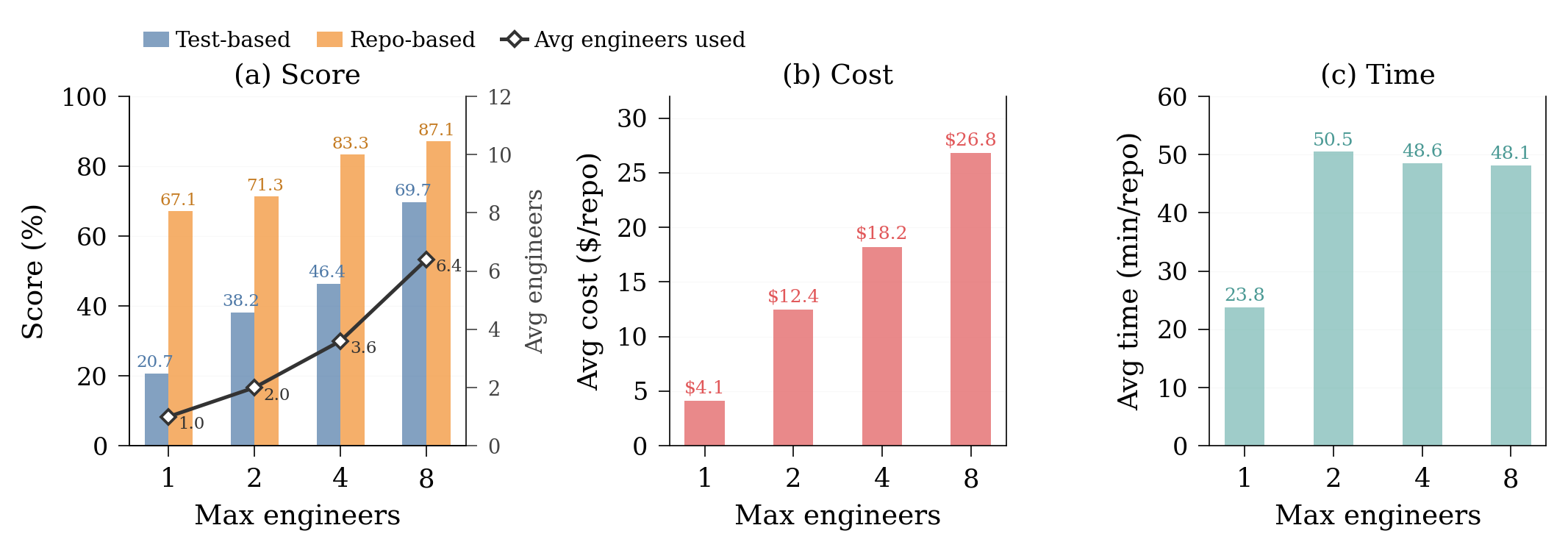}
  \caption{Scaling engineers with Sonnet~4.6 on Commit0-Lite. (a)~Both test-based and repo-based scores improve monotonically from 2 to 8 max engineers, and the line shows the average number actually deployed. (b)~Cost scales linearly with engineer count. (c)~Wall-clock time remains roughly constant due to parallel execution.}
  \label{fig:scaling}
\end{figure}

\subsection{Scaling to More Engineers}

A common finding in prior multi-agent systems is that performance \emph{degrades} as more agents are added~\citep{caid,cursorScalingAgents}. The reason is straightforward: for a fixed-size task, splitting work among more agents means each agent's sub-task becomes smaller, but the file coupling between sub-tasks increases. More agents need to touch shared interfaces, read overlapping files, and make mutually consistent design choices. Under worktree isolation, this coupling manifests as merge conflicts that grow combinatorially with the number of branches. STORM inherenetly avoid it given conflicts are detected and resolved at write time, higher coupling leads to more frequent but individually cheap rejections rather than a catastrophic merge failure at the end. Each conflict is resolved in isolation while the agent's reasoning context is still fresh, so scaling up agents does not increase the difficulty of final integration.

Figure~\ref{fig:scaling} shows the effect of increasing the maximum number of engineers from 2 to 8 on Sonnet~4.6 with STORM. Both the test-based score (overall pass rate) and the repo-based score (macro-average per-repository pass rate) improve: from 38.2\% to 69.7\% (+31.5) and 71.3\% to 87.1\% (+15.8), respectively. The gains are not uniform across the two transitions. Moving from 2 to 4 engineers yields +12.0 macro points but only +8.2 overall points, because the improvement concentrates on medium-sized repositories (\texttt{cookiecutter} 40.9$\to$98.6, \texttt{imapclient} 16.5$\to$89.1, \texttt{jinja} 0.0$\to$47.7). Moving from 4 to 8 engineers yields a smaller macro gain (+3.8) but a much larger overall gain (+23.3), driven almost entirely by \texttt{babel} (20.2$\to$57.5) and \texttt{jinja} (47.7$\to$66.5), repositories with thousands of tests and deep cross-file dependencies that benefit from aggressive parallelization.

Notably, the manager does not always use all available engineers. The average number of engineers actually deployed is 2.0, 3.6, and 6.4 for max settings of 2, 4, and 8 respectively. Simple repositories (\texttt{cachetools}, \texttt{deprecated}, \texttt{wcwidth}) consistently receive only 2 engineers regardless of the maximum, avoiding unnecessary coordination overhead. The number of repositories solved at $\geq$99\% pass rate grows from 7 (max=2) to 8 (max=4) to 10 (max=8) out of 16.

Cost scales approximately linearly with the number of engineers (\$199$\to$\$292$\to$\$429), while wall-clock time remains roughly constant ($\sim$13 hours total across all 16 repositories) because engineer tasks execute in parallel. The cost-efficiency ratio (dollars per percentage point of overall score) is stable at \$5--6 per point across all configurations, indicating that additional engineers provide proportional value rather than diminishing returns at this scale.

\subsection{Isolation Strategy}

\begin{table}[h]
  \caption{Effect of isolation strategy on Commit0-Lite with Claude Sonnet 4.6.}
  \label{tab:isolation-ablation}
  \centering
  \small
  \begin{tabular}{l cc cc}
    \toprule
Method & Score\textsubscript{w} $\uparrow$ & Score $\uparrow$ & Cost\textsubscript{eff} $\downarrow$ & Time\textsubscript{eff} $\downarrow$ \\
    \midrule
    Single-Agent           & 20.7 & \underline{66.4} & \textbf{3.2} & \underline{18.8} \\
    Soft Isolation         & 24.0 & 65.4 & 7.7 & 29.1 \\
    GitWorktree            & \underline{24.6} & 63.8 & 8.6 & 20.9 \\
    STORM                  & \textbf{46.2} & \textbf{82.5} & \underline{6.3} & \textbf{16.8} \\
    \bottomrule
  \end{tabular}
\end{table}

In Table~\ref{tab:isolation-ablation}, we compare isolation strategies on Commit0-Lite with Claude Sonnet 4.6 using 4 engineers. Prompt based soft isolation and git worktree isolation achieve similar results (Score\textsubscript{w} 24.0 vs.\ 24.6, Score 65.4 vs.\ 63.8), both improving over the single-agent baseline in weighted score (20.7) while remaining comparable in unweighted score. This suggests that delegation alone provides gains on larger repositories where multiple agents can cover more test cases in parallel, but the choice between instruction-level constraints and physical branch separation has limited impact on final pass rate. STORM outperforms both isolation strategies (Score\textsubscript{w} 46.2, Score 82.5) despite using the same number of engineers. First, STORM detects conflicts at write time rather than deferring them: when an edit violates local state consistency (Eq.~\ref{eq:validity}), it is rejected immediately, letting the agent re-plan while its reasoning context is still fresh. Soft isolation instead relies on instruction-level constraints that are frequently violated, causing silent overwrites, while worktree isolation defers conflicts to a merge step where multi-file resolution often fails. Second, because all engineers share a single workspace, any read returns the latest committed state, including other engineers' recent changes. Agents working on adjacent modules naturally observe each other's implementations and can adapt interfaces accordingly, whereas worktree isolation keeps engineers blind to concurrent progress until merge time, by which point incompatible decisions may have already propagated.

\begin{figure}[h]
    \centering
    \includegraphics[width=\linewidth]{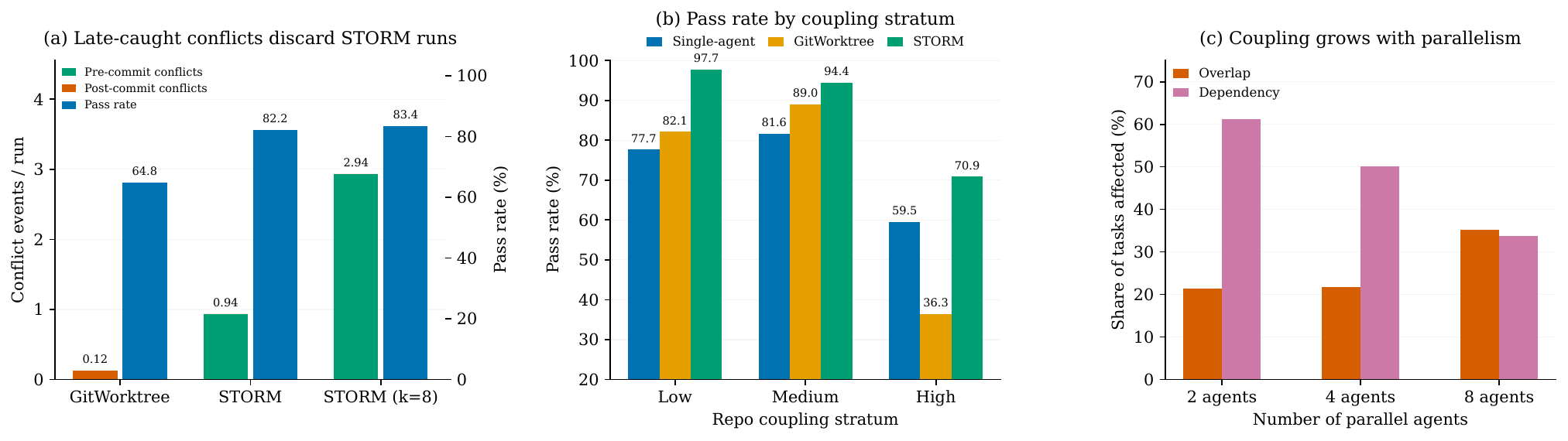}
    \caption{Analysis summary on Sonnet~4.6. (a) STORM surfaces conflicts pre-commit, while GitWorktree leaves most conflicts for late (post-commit/merge) resolution; late-caught conflicts are associated with lower final pass rates (blue overlay). (b) STORM's pass-rate advantage over the single-agent and GitWorktree baselines grows with cross-file coupling. (c) Narrow task scopes are not always independent scopes: first-round overlap and dependency signals remain common, and rise with $k$. Quantities in (a) and (c) are proxy measurements derived from manager-review events and declared task scopes.}
    \label{fig:analysis-three}
  \end{figure}

\subsection{Further Analysis}
\label{sec:analysis}

\paragraph{STORM helps by exposing invalid parallel work early.}
  STORM's benefit is converting hidden integration errors into explicit write-time feedback. As Figure~\ref{fig:analysis-three}(a) shows, on Sonnet~4.6 GitWorktree leaves most conflict events until post-commit merge while STORM surfaces them pre-commit at both $k{=}4$ and $k{=}8$, and the configurations that catch conflicts early also achieve higher final pass rates. The effect is not that STORM eliminates conflicts, but that it relocates them from a fragile post-hoc merge to a cheap write-time check. GitWorktree produces 3.81 reviewed write attempts per run at 91.8\% acceptance, versus 6.00 at 81.2\% for STORM ($k{=}4$) and 10.25 at 67.1\% for STORM ($k{=}8$); the lower acceptance reflects a stricter consistency filter rather than weaker engineering, and STORM still outperforms GitWorktree in final repository quality.

  \paragraph{STORM remains robust on high-coupling repositories where alternatives collapse.}
  STORM's advantage is sharpest where parallel coordination matters most. Stratifying repositories by a proxy coupling score (from task overlap, dependency signals, multi-file scope, and rejection evidence), STORM's lead over GitWorktree widens from $+15.6$ points on low-coupling and $+5.4$ on medium-coupling subsets to $+34.6$ points on the high-coupling stratum (Figure~\ref{fig:analysis-three}(b)), because GitWorktree collapses outright there ($36.3\%$ vs.\ STORM $70.9\%$). STORM's margin over the single-agent baseline remains positive across all strata ($+20.0$, $+12.8$, and $+11.4$ points from low to high); it narrows in the high-coupling regime only because the absolute ceiling falls for every method, not because STORM loses its edge. Concretely, as coupling scales from low to high, STORM degrades gracefully ($97.7\% \to 94.4\% \to 70.9\%$), whereas single-agent falls from $77.7\%$ to $59.5\%$ and GitWorktree plummets from $82.1\%$ and $89.0\%$ to $36.3\%$: STORM is the \emph{only} method that does not break down under coupling. The repository-level pattern agrees: \texttt{marshmallow} rises $0.0\%\to82.3\%$ and \texttt{imapclient} $9.7\%\to89.1\%$ (Table~\ref{tab:per-repo}), precisely where branch-level isolation is most brittle. As a result, STORM more than doubles the Sonnet weighted score of the single agent ($46.2$ vs.\ $20.7$) and nearly doubles that of GitWorktree ($46.2$ vs.\ $24.6$).

  \paragraph{Scaling is limited more by decomposition quality than by STORM itself.}
  The diminishing returns of additional engineers come from decomposition, not STORM. As Figure~\ref{fig:analysis-three}(c) illustrates, STORM already produces narrow scopes: 91.4\% of $k{=}4$ tasks are single-file, yet 50.0\% still carry dependency signals and 21.7\% of first-round tasks overlap another's file scope. At $k{=}8$, single-file share stays high (85.1\%) while first-round overlap rises to 35.1\%. First-round overlap correlates with rejected-review rate ($r=0.28$ overall, $r=0.78$ within $k{=}8$): once the manager can no longer partition the repository into disjoint units, additional engineers create contention faster than useful parallel work.

\begin{figure}[t]
  \centering

  \begin{subfigure}[t]{\linewidth}
    \centering
    \includegraphics[width=\linewidth]{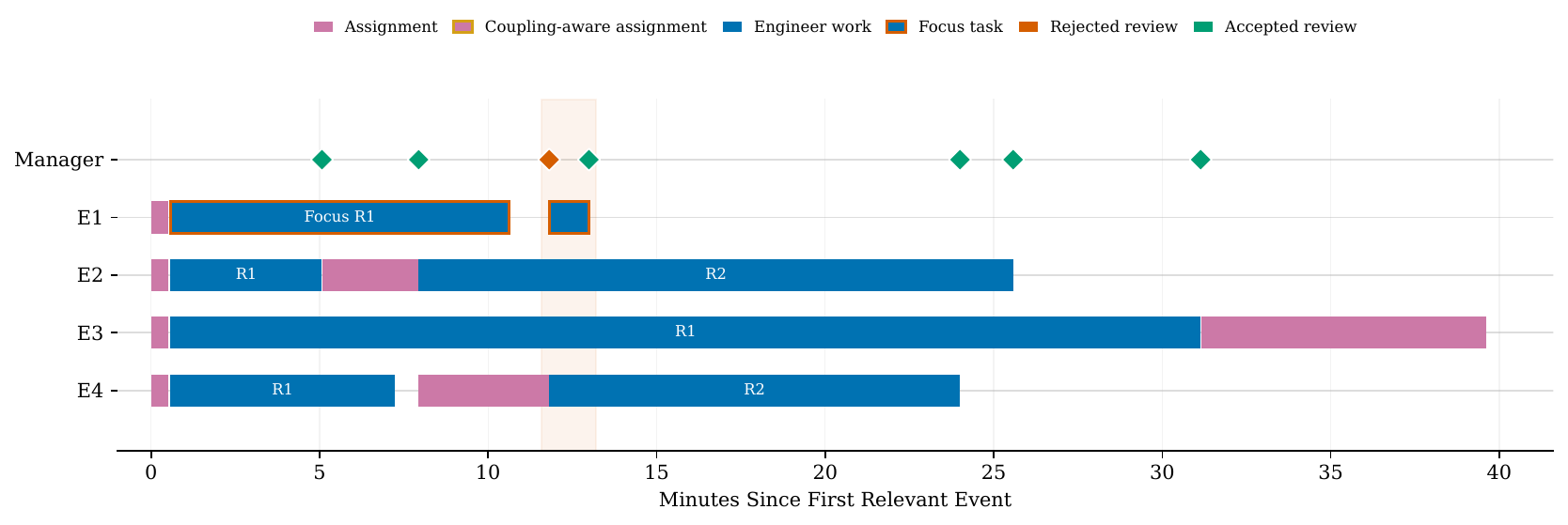}
    \caption{GitWorktree: post-hoc detection.}
    \label{fig:gantt-pre-vs-post-a}
  \end{subfigure}
  
  \begin{subfigure}[t]{\linewidth}
    \centering
    \includegraphics[width=\linewidth]{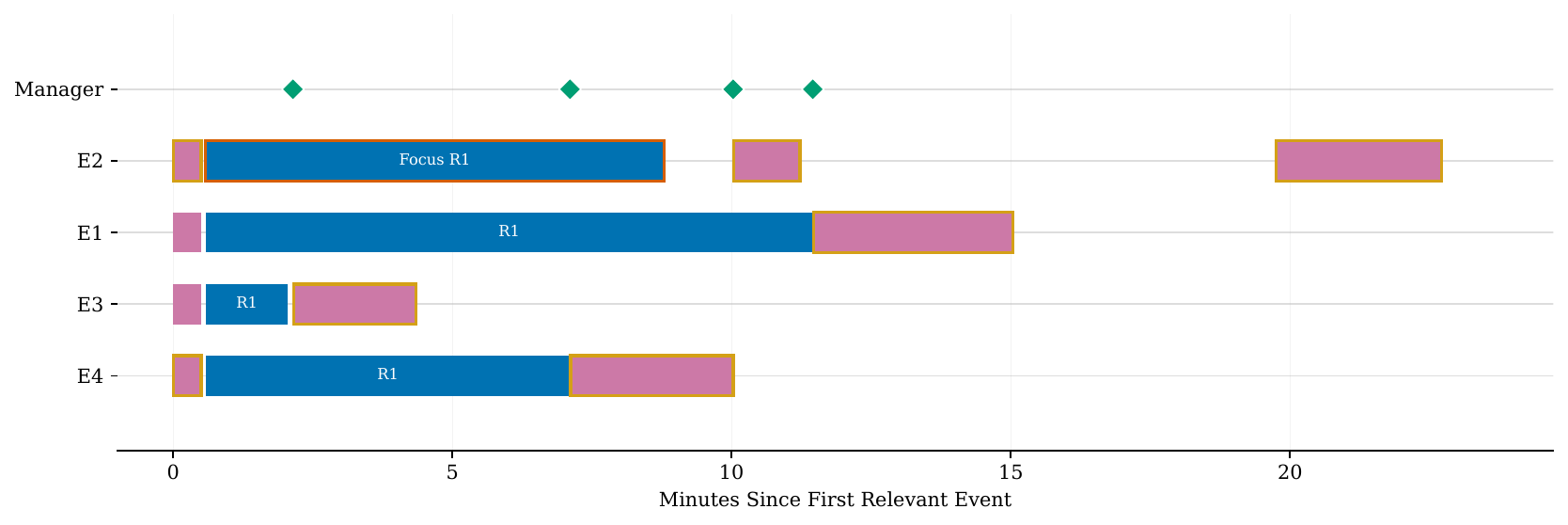}
    \caption{STORM: pre-hoc coordination.}
    \label{fig:gantt-pre-vs-post-b}
  \end{subfigure}
  
  \caption{Paired run timelines on \texttt{jinja}; the legend in (a) applies to both panels. (a) GitWorktree exposes the \texttt{utils.py} coupling only at merge review: the red diamond marks the rejected review, the shaded band the rework window, and \emph{Focus~R2} the retry that is eventually accepted. (b) STORM detects the same coupling at decomposition time and co-assigns or sequences the dependent tasks (gold-outlined manager instructions carry explicit interdependence reasoning); the focus task passes on its first review and no rejection remains within the coupled task set.}
  \label{fig:gantt-pre-vs-post}
\end{figure}

  \paragraph{Case study: pre-hoc coordination versus post-hoc recovery.}
  A paired run on \texttt{jinja} makes the aggregate pattern in Figure~\ref{fig:analysis-three}(a) concrete at the level of a single repository. Under GitWorktree (Figure~\ref{fig:gantt-pre-vs-post-a}), four engineers work on isolated branches and the coupling around \texttt{utils.py} stays invisible until merge: the manager rejects the focus diff with an explicit \emph{merge conflict} reason (red diamond) and the same engineer reworks the task in a second round inside the shaded recovery window before acceptance. Under STORM (Figure~\ref{fig:gantt-pre-vs-post-b}, restricted to the coupling task set), the manager reasons about the same coupling at decomposition and instruction time: it packs \texttt{utils.py} and \texttt{async\_utils.py} into a single assignment and explicitly sequences downstream consumers against it (gold-outlined bars carry this interdependence reasoning); the focus task then passes on the first review and no rejection appears in the coupled task set. The two timelines describe the same mechanism from opposite sides, GitWorktree lets coupling surface \emph{post hoc} at the merge boundary, while STORM converts the identical signal into a \emph{pre hoc} coordination decision baked into how tasks are defined and ordered.

In summary, the analysis suggests a clear interpretation of STORM. Its main contribution is not cheaper computation or perfect recovery from every rejected edit. Its contribution is to replace delayed, fragile integration with immediate consistency checks that make parallel collaboration substantially more reliable on the repositories where parallelism is most useful. The residual failure modes of STORM are further analyzed in Appendix~\ref{app:analysis-failure-details}.

\section{Related Work}
\label{sec:related}

\paragraph{Multi-Agent Collaboration}
\label{sec:related-coll}

Multi-agent collaboration decomposes complex tasks into specialized roles~\citep{selfcollaboration,chatdev,metagpt,autogen}. Compared with a single-agent workflow, a multi-agent system can separate planning, implementation, testing, debugging, and review~\citep{magis,swedebate,agentforge}. Existing work mainly studies role-based software development, repository-level issue resolution, and workspace-isolated parallel development. Self-Collaboration \citep{selfcollaboration} first introduced the multi-agent framework for code generation \citep{codeagentsurvey}\nocite{dong2026iocodediscoveryagent}. Subsequently, ChatDev and MetaGPT model software development as a structured workflow among specialized agents~\citep{chatdev,metagpt}. MAGIS and SWE-Debate study multi-agent issue resolution on SWE-style benchmarks~\citep{magis,swedebate}. AgentForge adds execution feedback through Docker-based validation~\citep{agentforge}. Git worktree is a common mechanism for workspace-isolated parallel agents. Systems such as Git Worktree, CORAL, and StatsClaw use git worktree to separate agent workspaces~\citep{caid,coral,statsclaw}. Each agent works in a separate branch or worktree, while shared memory or a central coordinator supports information exchange. Workspace isolation enables parallel exploration and reduces direct interference between agents' code changes.
However, git worktree only provides low-level workspace isolation. It does not solve task decomposition, dependency tracking, semantic conflicts, or merge selection. Different agents may edit related files under incompatible assumptions. Some errors may only appear after integration. Industry systems also report that optimistic concurrency control improves over lock-based shared-state coordination~\citep{cursorScalingAgents}, but remains insufficient without hierarchical task decomposition. Therefore, worktree-based systems still require higher-level coordination, shared memory, review, and execution-based verification.

\paragraph{State Management in Agentic Systems}
\label{sec:related-state}

Agentic systems require persistent state across multi-step interaction. The state may include conversation history, plans, tool outputs, execution feedback, repository changes, and intermediate artifacts. Existing work manages agent state through reflection, memory, skill libraries, and execution traces~\citep{reflexion,generativeagents,memgpt,voyager,hiagent}. Reflexion and Generative Agents use reflection and episodic memory to improve later decisions~\citep{reflexion,generativeagents}. MemGPT manages long-term context through memory tiers~\citep{memgpt}. Voyager stores reusable skills as executable code, while HiAgent manages working memory with subgoals~\citep{voyager,hiagent}. However, state management remains difficult in multi-agent software development. Most existing methods focus on a single agent or a single memory store. Multi-agent systems must also manage local states, shared states, workspace states, and execution states. Git Worktree uses isolated workspaces and branch-and-merge coordination for asynchronous software engineering agents~\citep{caid}. CORAL and StatsClaw use shared memory and isolated worktrees to support parallel execution~\citep{coral,statsclaw}. CodeCRDT instead coordinates agents through observable shared state with deterministic convergence~\citep{codecrdt}. However, workspace isolation and shared memory do not guarantee state consistency, dependency tracking, or safe integration. Reliable multi-agent software development still requires explicit control over what state is stored, shared, updated, and used.

\section{Conclusion}
\label{sec:conclusion}

We presented STORM, a state management framework that replaces workspace isolation with local state consistency for multi-agent collaboration. All agents share one workspace; writes are accepted only if the agent's observed files remain unchanged, and intent annotations coordinate semantics at shared boundaries. On Commit0-Lite, STORM achieves 82.5\% macro / 46.2\% weighted pass rate with Sonnet~4.6 (vs.\ 66.4\% / 20.7\% single-agent, 63.8\% / 24.6\% GitWorktree). On PaperBench Code-Dev, STORM scores 74.1 vs.\ 72.7 and 68.7. Results generalize across Qwen and DeepSeek, with STORM-Combined reaching 88.2 / 76.2 on Qwen. Scaling to 8 engineers improves score monotonically with constant wall-clock time. These results suggest that explicit state management is a more effective foundation for multi-agent collaboration than workspace isolation.

\bibliographystyle{unsrtnat}
\bibliography{ref}

\newpage
\appendix

\section{Detailed Experiment Setup}
\label{appendix:setup}

We evaluate on \textbf{Commit0}~\citep{commit0}, a repository-level code implementation benchmark. Each instance is a real Python repository with its test suite intact but source implementations removed (replaced with stubs or \texttt{pass} statements). The task is to implement the missing code so that the tests pass. We use the Commit0 Lite subset: 16 repositories spanning a range of domains and sizes. Several repositories require cross-file reasoning, making them difficult for a single agent and a natural fit for studying multi-agent collaboration.
We also evaluate on \textbf{PaperBench}~\citep{PaperBench}, a benchmark for evaluating AI agents on end-to-end research replication. Each instance is based on an ICML 2024 Spotlight or Oral paper. The task is to understand the paper, implement the required codebase from scratch, run experiments, and reproduce the key results. PaperBench uses hierarchical rubrics to evaluate fine-grained replication progress rather than only final execution success. These tasks require long-horizon planning, cross-file implementation, experiment execution, and result verification, making them a natural fit for studying multi-agent collaboration.

\paragraph{Configurations.} We compare five configurations across three LLMs (Claude Sonnet 4.6, Qwen 3.6 Plus, DeepSeek V4 Pro): (1) \textbf{Single-agent} with 100 LLM iterations; (2) \textbf{GitWorktree}~\citep{caid} with 4 engineers in isolated worktrees, merged after completion; (3) \textbf{STORM} with a manager and 4 engineers sharing one workspace; (4) \textbf{GitWorktree-Combined}; and (5) \textbf{STORM-Combined}. The Combined variants run both single-agent and multi-agent, keeping the per-test union of passed tests, but only invoking multi-agent on repositories where single-agent did not already achieve 100\%.

\paragraph{Implementation.} All agents run on the OpenHands SDK~\citep{openhands} inside Docker containers. The manager gets up to 50 LLM iterations; each engineer gets up to 80 per task and can be reassigned once. During analysis, the manager recommends how many engineers to use based on repository complexity; for simple repositories it may use as few as 2. Evaluation uses \texttt{pytest} with \texttt{-{}-json-report}; if the JSON report is unavailable we fall back to parsing terminal output. The total number of tests per repository is taken from the ground-truth counts in the Commit0 paper~\citep{commit0}.

\paragraph{Metrics.} We report \textbf{Score\textsubscript{w}} (weighted pass rate: total tests passed / total tests, dominated by large repositories) and \textbf{Score} (macro: mean of per-repository pass rates). For efficiency we report \textbf{Cost\textsubscript{eff}} (total dollars / Score\textsubscript{w}) and \textbf{Time\textsubscript{eff}} (total minutes / Score\textsubscript{w}), both lower-is-better.

\paragraph{PaperBench.} We use the PaperBench Code-Dev subset, which evaluates code development only and skips the reproduction execution step. The judge grades only ``Code Development'' nodes in the rubric. We use Sonnet~4.6 as the LLM judge. Due to the high per-run cost (20 papers per configuration), we report PaperBench results on Sonnet~4.6 agents.

\section{Prompt templates}
\label{app:prompts}

We summarize the key prompts used in the manager-engineer protocol. The full prompt text is available in the released code.

\begin{center}
{\setlength{\fboxsep}{8pt}
\fcolorbox{orange!70!black}{orange!3}{%
\begin{minipage}{0.92\linewidth}
\small
\textbf{Manager scan instruction (excerpt):}\\[4pt]
Start by exploring the repository structure to understand what your team needs to work on and identify which functions have top priority to be implemented (i.e., those with \texttt{pass} statements). To explore the repository structure and dependencies:\\[4pt]
1. Check the imports and the actual functions used in the files with \texttt{pass} statements and review the relevant tests to understand the EXPECTED BEHAVIOR of the functions and the dependencies between the files.\\[2pt]
2. Run \texttt{pytest --collect-only --continue-on-collection-errors} to have a better understanding of the scope and the dependencies of this codebase.\\[4pt]
Collect all the undefined functions from the test collection errors and add them with clear docstrings and \texttt{pass} statements into the files and make a local commit. DO NOT make any changes to the functions with \texttt{pass} statements since those need to be implemented by the engineers.
\end{minipage}}}
\end{center}

\paragraph{Manager delegation.}

\begin{center}
{\setlength{\fboxsep}{8pt}
\fcolorbox{orange!70!black}{orange!3}{%
\begin{minipage}{0.92\linewidth}
\small
\textbf{Manager delegation instruction (excerpt):}\\[4pt]
Split the overall implementation work into up to $k$ major tasks, balancing complexity and estimated effort as evenly as possible. Make sure the high dependent files are in the same major task. Try to split the major tasks at file level first; if a single file contains a disproportionately large amount of functions with \texttt{pass} statements, you can delegate at the function level and assign non-overlapping sets of functions to multiple engineers.\\[4pt]
For each engineer, assign the first task that has the highest priority within their major task. When you provide instruction to each engineer, briefly summarize the relevant repository structure and the dependencies so they don't need to re-explore the repository. Then clearly specify which file/functions to implement and explain the purpose of this task. If the assigned functions depend on other stub functions, include a brief description of what each dependency does.
\end{minipage}}}
\end{center}

\paragraph{Engineer task prompt.} Each engineer receives the repository structure summary, its assigned functions, dependency descriptions, the test command, and shared-workspace rules. The key constraint is shown below:

\begin{center}
{\setlength{\fboxsep}{8pt}
\fcolorbox{orange!70!black}{orange!3}{%
\begin{minipage}{0.92\linewidth}
\small
\textbf{Engineer task prompt (excerpt):}\\[4pt]
You are a software engineer working on implementing a python code repository in a group. You are responsible for implementing the functions instructed by your manager (i.e., the functions with \texttt{pass} statements) and passing the unit tests.\\[4pt]
\textbf{Shared Workspace:} Your workspace is \texttt{\{worktree\_path\}}. ALL engineers on your team share this SAME directory. DO NOT modify files that belong to other engineers. Only edit the specific files and functions assigned to you. Check before modifying any file to make sure another engineer hasn't already changed it.\\[4pt]
\textbf{Scope Discipline:} You are ONLY responsible for implementing the functions listed below. If you see test failures caused by functions in OTHER files or functions NOT assigned to you, DO NOT attempt to fix them. Report the failure to your manager and move on.\\[4pt]
\textbf{Multi-agent coordination:} Every edit you make MUST include a one-line comment in the form \texttt{\# \{engineer\_id\}: <short intent>} IMMEDIATELY above the block you added or changed. If you see \texttt{\# <other-engineer-id>: ...} comments, preserve both the comment and the block below it unless your task explicitly requires changing them.\\[4pt]
You are assigned to implement the following functions in the file: \texttt{\{file\_path\}}: \texttt{\{functions\}}. After you finished the implementation, make sure it will not cause any hanging issues. Do NOT commit; the manager will review your changes and commit on your behalf.
\end{minipage}}}
\end{center}

\paragraph{Manager review and reassignment.} After each round, the manager checks commit status. For successful commits, it assigns the next task. For failed commits, it reassigns the same task. In the final review, the manager merges all engineers' work, resolves integration issues (import mismatches, naming inconsistencies), and checks for hanging code (infinite loops, \texttt{input()} calls).

\begin{center}
{\setlength{\fboxsep}{8pt}
\fcolorbox{orange!70!black}{orange!3}{%
\begin{minipage}{0.92\linewidth}
\small
\textbf{Manager delegation output format (excerpt):}\\[4pt]
\texttt{\{}\\
\quad\texttt{"engineer\_id": "engineer\_1",}\\
\quad\texttt{"file\_path": "path/to/file.py",}\\
\quad\texttt{"functions\_to\_implement": ["func1", "func2"],}\\
\quad\texttt{"instruction": "Implement the tensor storage}\\
\quad\quad\texttt{layout. The TensorData class uses ..."}\\
\texttt{\}}
\end{minipage}}}
\end{center}

\begin{table}[t!]
  \caption{Per-repository pass rate, cost, and time (seconds) for Claude Sonnet~4.6. Best pass rate per row in bold.}
  \label{tab:per-repo}
  \centering
  \small
  \begin{tabular}{l rrr rrr rrr}
    \toprule
    & \multicolumn{3}{c}{\textbf{Single-Agent}} & \multicolumn{3}{c}{\textbf{GitWorktree (4 Eng)}} & \multicolumn{3}{c}{\textbf{STORM (4 Eng)}} \\
    \cmidrule(lr){2-4} \cmidrule(lr){5-7} \cmidrule(lr){8-10}
    Repository & Rate & Cost & Time & Rate & Cost & Time & Rate & Cost & Time \\
    \midrule
    babel        &   0.6 &  4.3 &  934 &   1.3 & 30.8 & 3301 & \textbf{20.2} &  7.9 & 1678 \\
    cachetools   & \textbf{100.0} &  0.7 &  249 & \textbf{100.0} &  2.6 &  657 & \textbf{100.0} &  1.5 &  400 \\
    chardet      & \textbf{99.7} &  7.0 &  768 &   0.3 &  3.2 &  817 &  22.1 & 18.7 & 2932 \\
    cookiecutter &  74.9 &  7.2 & 1958 &  94.3 & 24.2 & 2548 & \textbf{98.6} & 24.8 & 5234 \\
    deprecated   & \textbf{100.0} &  1.1 &  698 &  94.7 &  1.4 &  528 & \textbf{100.0} &  3.4 &  898 \\
    imapclient   &   9.7 &  2.6 & 3141 &  37.8 &  6.3 & 1698 & \textbf{89.1} & 24.9 & 3736 \\
    jinja        &   0.0 &  4.7 &  665 &   5.8 & 29.2 & 3332 & \textbf{47.1} & 39.3 & 3733 \\
    marshmallow  &   0.0 &  5.9 & 1150 &  60.9 & 31.7 & 3162 & \textbf{82.3} & 32.8 & 3896 \\
    minitorch    & \textbf{70.4} &  6.8 & 2093 &  53.0 & 16.1 & 2337 & \textbf{70.4} & 35.9 & 4439 \\
    parsel       & \textbf{100.0} &  6.3 & 1568 &  93.2 &  4.3 & 1202 & \textbf{100.0} & 11.0 & 1974 \\
    portalocker  & \textbf{100.0} &  7.2 & 2273 & \textbf{100.0} &  8.8 & 1741 & \textbf{100.0} & 11.1 & 1727 \\
    pyjwt        & \textbf{100.0} &  3.5 &  716 &   5.8 &  2.2 &  639 & \textbf{100.0} & 18.6 & 1783 \\
    simpy        &  58.6 &  0.5 & 2109 &  78.6 &  6.2 & 1627 & \textbf{100.0} & 21.9 & 2655 \\
    tinydb       & \textbf{99.5} &  2.5 &  940 & \textbf{99.5} & 10.3 & 1353 & \textbf{99.5} & 11.1 & 1474 \\
    voluptuous   &  49.7 &  4.2 & 3621 & \textbf{96.0} & 32.3 & 4928 &  90.6 & 25.9 & 9181 \\
    wcwidth      & \textbf{100.0} &  1.5 &  442 & \textbf{100.0} &  2.4 &  956 & \textbf{100.0} &  2.9 &  869 \\
    \bottomrule
  \end{tabular}
\end{table}

\begin{table}[t!]
  \caption{Per-repository pass rate, cost, and time (seconds) for Qwen 3.6 Plus. Best pass rate per row in bold.}
  \label{tab:per-repo-qwen36plus}
  \centering
  \small
  \begin{tabular}{l rrr rrr rrr}
    \toprule
    & \multicolumn{3}{c}{\textbf{Single-Agent}} & \multicolumn{3}{c}{\textbf{GitWorktree (4 Eng)}} & \multicolumn{3}{c}{\textbf{STORM (4 Eng)}} \\
    \cmidrule(lr){2-4} \cmidrule(lr){5-7} \cmidrule(lr){8-10}
    Repository & Rate & Cost & Time & Rate & Cost & Time & Rate & Cost & Time \\
    \midrule
    babel        &   3.6 &  3.7 & 1709 &   0.2 &  2.0 &  557 & \textbf{74.2} & 30.1 & 5502 \\
    cachetools   & \textbf{100.0} &  0.6 &  487 & \textbf{100.0} &  3.2 &  966 & \textbf{100.0} &  2.2 &  927 \\
    chardet      & \textbf{99.7} &  3.0 & 1159 & \textbf{99.7} &  3.2 &  966 &  38.8 & 12.1 & 2350 \\
    cookiecutter &  38.7 &  3.0 & 1485 &  36.5 &  8.6 & 4234 & \textbf{56.4} &  9.2 & 3733 \\
    deprecated   & \textbf{100.0} &  1.0 & 1135 & \textbf{100.0} &  3.8 & 2369 & \textbf{100.0} &  2.1 & 1124 \\
    imapclient   & \textbf{100.0} &  2.3 & 1193 &   0.0 & 12.1 & 4801 &  44.6 & 19.2 & 4441 \\
    jinja        & \textbf{98.8} &  0.6 &  366 &   0.0 &  2.6 &  744 &   0.0 &  2.6 &  744 \\
    marshmallow  & \textbf{28.5} &  3.9 & 2714 &   0.0 & 15.1 & 3106 &  23.7 &  7.7 & 2386 \\
    minitorch    &  33.9 &  4.3 & 1664 & \textbf{52.2} &  8.0 & 2463 &  30.4 &  5.4 & 1551 \\
    parsel       & \textbf{100.0} &  1.4 &  591 &   6.3 &  1.1 &  546 &  82.0 & 10.7 & 3239 \\
    portalocker  &  83.3 &  0.7 & 3623 & \textbf{100.0} &  4.0 & 8099 & \textbf{100.0} &  5.9 & 4074 \\
    pyjwt        &  97.3 &  4.3 & 1866 &  89.2 & 14.8 & 6453 & \textbf{100.0} & 13.9 & 3814 \\
    simpy        &  76.4 &  3.8 & 3132 &  51.4 & 12.8 & 8926 & \textbf{81.4} &  8.7 & 3528 \\
    tinydb       &  88.1 &  3.0 & 1365 & \textbf{99.5} &  9.4 & 2940 &  96.0 &  7.0 & 3550 \\
    voluptuous   &  55.7 &  4.5 & 1346 &  83.2 & 10.9 & 2491 & \textbf{100.0} & 12.2 & 2685 \\
    wcwidth      & \textbf{100.0} &  3.9 & 2599 & \textbf{100.0} &  4.5 & 6393 & \textbf{100.0} &  4.0 & 2544 \\
    \bottomrule
  \end{tabular}
\end{table}

\section{Full experimental results}
\label{app:full-results}

Tables~\ref{tab:per-repo}--\ref{tab:sonnet-paperbench} report per-instance results for all models. We highlight several patterns.

\paragraph{STORM excels on cross-file repositories.} The largest gains over both baselines come from repositories with heavy inter-file dependencies. On Sonnet, \texttt{marshmallow} jumps from 0.0\% (single) and 60.9\% (GitWorktree) to \textbf{82.3\%} under STORM; \texttt{imapclient} from 9.7\% / 37.8\% to \textbf{89.1\%}; and \texttt{jinja} from 0.0\% / 5.8\% to \textbf{47.1\%}. On Qwen, \texttt{babel} shows the most dramatic improvement: 3.6\% (single) and 0.2\% (GitWorktree) to \textbf{74.2\%} under STORM, demonstrating that shared-workspace coordination is critical for large, tightly coupled codebases.

\paragraph{Single-agent strengths.} The single agent remains competitive on small, self-contained repositories (\texttt{chardet}, \texttt{deprecated}, \texttt{parsel}) where decomposition overhead outweighs parallelism benefits. On Sonnet, \texttt{chardet} scores 99.7\% single vs.\ 22.1\% STORM, showing that STORM's decomposition can occasionally hurt when the repository is better solved monolithically.

\paragraph{PaperBench patterns.} On PaperBench Code-Dev (Table~\ref{tab:sonnet-paperbench}), STORM leads on 11 of 20 papers, GitWorktree on 6, and single-agent on 3. STORM's largest wins come on papers requiring substantial code organization (\texttt{what-will-my-model-forget}: 99.8 vs.\ 82.9 single, \texttt{lbcs}: 95.6 vs.\ 84.1 single). GitWorktree wins on papers where independent sub-tasks map cleanly to separate files (\texttt{sample-specific-masks}: 98.2 vs.\ 72.8 STORM), suggesting that when task boundaries align perfectly with file boundaries, isolation incurs no penalty.

\begin{table}[t!]
  \caption{Per-repository pass rate, cost, and time (seconds) for DeepSeek V4 Pro. Best pass rate per row in bold.}
  \label{tab:per-repo-deepseek}
  \centering
  \small
  \begin{tabular}{l rrr rrr rrr}
    \toprule
    & \multicolumn{3}{c}{\textbf{Single-Agent}} & \multicolumn{3}{c}{\textbf{GitWorktree (4 Eng)}} & \multicolumn{3}{c}{\textbf{STORM (4 Eng)}} \\
    \cmidrule(lr){2-4} \cmidrule(lr){5-7} \cmidrule(lr){8-10}
    Repository & Rate & Cost & Time & Rate & Cost & Time & Rate & Cost & Time \\
    \midrule
    babel        &   1.2 &  3.9 & 3078 &   0.7 &  7.3 & 6153 & \textbf{19.3} &  4.9 & 6344 \\
    cachetools   &  71.2 &  0.1 & 1275 & \textbf{100.0} &  1.1 & 2016 & \textbf{100.0} &  1.3 & 1561 \\
    chardet      & \textbf{99.7} &  2.6 & 2044 &   0.3 &  0.9 &  987 &   1.3 &  3.9 & 3824 \\
    cookiecutter & \textbf{69.5} &  4.3 & 3627 &  34.3 &  4.1 & 2927 &  35.1 &  3.0 & 2569 \\
    deprecated   & \textbf{100.0} &  0.5 & 3308 &   9.9 &  0.3 &  399 & \textbf{100.0} &  1.6 & 2148 \\
    imapclient   &  16.9 &  3.5 & 3622 &   9.7 &  3.2 & 4337 & \textbf{49.1} &  6.4 & 6320 \\
    jinja        & \textbf{29.3} &  5.0 & 3347 &   3.3 & 13.6 & 6531 &   5.8 & 11.7 & 6778 \\
    marshmallow  &  38.6 &  5.2 & 3282 & \textbf{49.0} & 12.8 & 4745 &  47.9 &  9.3 & 5402 \\
    minitorch    &  36.1 &  3.4 & 3636 & \textbf{51.7} &  3.9 & 2613 &  44.3 &  4.4 & 3018 \\
    parsel       &  97.6 &  3.3 & 2861 & \textbf{100.0} &  4.7 & 3434 & \textbf{100.0} & 12.5 & 8645 \\
    portalocker  &  36.1 &  0.1 & 3621 & \textbf{97.2} &  2.5 & 5894 & \textbf{97.2} &  3.1 & 3708 \\
    pyjwt        &  88.8 &  4.4 & 3625 & \textbf{91.9} &  4.2 & 2214 &  81.1 &  6.3 & 6668 \\
    simpy        & \textbf{73.6} &  3.2 & 3625 &  52.9 &  6.6 & 4507 &  66.4 &  6.9 & 4990 \\
    tinydb       &  99.5 &  2.4 & 2188 & \textbf{100.0} &  4.5 & 3086 &  96.5 &  4.2 & 4873 \\
    voluptuous   & \textbf{85.9} &  4.3 & 3621 &     0.0 &    - &    - &  67.8 & 15.2 & 6271 \\
    wcwidth      & \textbf{100.0} &  0.7 & 3513 &   2.6 &  0.3 &  277 & \textbf{100.0} &  2.1 & 2487 \\
    \bottomrule
  \end{tabular}
\end{table}

\begin{table}[h]
  \caption{Per-paper scores, cost, and wall-clock time for Claude Sonnet 4.6 on PaperBench Code-Dev. Best score per row in bold.}
  \label{tab:sonnet-paperbench}
  \centering
  \resizebox{\textwidth}{!}{
  \begin{tabular}{l rrr rrr rrr}
    \toprule
    & \multicolumn{3}{c}{\textbf{Single-Agent (100 iter)}} & \multicolumn{3}{c}{\textbf{GitWorktree (2 Eng)}} & \multicolumn{3}{c}{\textbf{STORM (2 Eng)}} \\
    \cmidrule(lr){2-4} \cmidrule(lr){5-7} \cmidrule(lr){8-10}
    Paper & Score & Cost & Time & Score & Cost & Time & Score & Cost & Time \\
    \midrule
    sample-specific-masks                         & 93.9 & 28.40 & 2721 &  \textbf{98.2} & 38.81 &  5555 & 72.8 & 22.64 & 6156 \\
    bam                                           & 97.4 &  5.34 & 1832 &  97.7 & 110.97 &  7067 & \textbf{97.9} & 14.27 & 3806 \\
    what-will-my-model-forget                     & 82.9 & 207.31 & 3601 &  94.7 & 215.28 &  3149 & \textbf{99.8} & 315.79 & 3937 \\
    pinn                                          & 81.0 & 42.22 & 3600 &  \textbf{91.1} & 20.61 &  6459 & 69.6 & 10.58 & 5328 \\
    sequential-neural-score-estimation            & \textbf{92.6} & 28.33 & 2060 &  88.7 & 35.29 &  4412 & 91.0 & 39.91 & 6699 \\
    mechanistic-understanding                     & \textbf{94.4} & 15.69 & 2558 &  81.9 & 17.66 &  4404 & 61.9 & 0.00 & 6246 \\
    test-time-model-adaptation                    & 58.5 &  5.75 & 2084 &  \textbf{81.4} & 36.87 &  3694 & 80.8 & 40.42 & 4666 \\
    robust-clip                                   & 51.1 & 21.03 & 1843 &  78.4 & 34.59 &  3677 & \textbf{82.6} & 45.45 & 4029 \\
    fre                                           & 72.5 & 68.11 & 1438 &  78.0 & 79.55 &  2971 & \textbf{78.4} & 20.49 & 5448 \\
    stochastic-interpolants                       & 78.4 & 13.66 & 1902 &  76.4 & 18.47 &  2423 & \textbf{79.3} & 22.63 & 4888 \\
    adaptive-pruning                              & 43.6 & 34.99 & 2863 &  72.7 & 45.88 &  6724 & \textbf{72.8} & 26.60 & 5688 \\
    all-in-one                                    & \textbf{87.3} & 27.88 & 2991 &  72.0 & 46.90 &  5894 & 72.6 & 20.66 & 6389 \\
    stay-on-topic-with-classifier-free-guidance   & 67.5 & 22.92 & 2651 &  \textbf{70.4} & 29.61 &  5604 & 69.2 & 38.31 & 4933 \\
    bbox                                          & 43.8 & 34.22 & 2061 &  \textbf{68.0} & 46.35 &  3438 & 50.5 & 12.44 & 5443 \\
    lca-on-the-line                               & 72.0 & 62.13 & 1791 &  66.6 & 147.78 &  5730 & \textbf{75.7} & 25.93 & 4899 \\
    ftrl                                          & 53.5 & 37.82 & 1965 &  63.4 & 51.09 &  5506 & \textbf{66.1} & 26.62 & 5302 \\
    rice                                          & 41.0 & 44.08 & 1934 &  59.7 & 76.31 &  7122 & \textbf{61.9} & 70.25 & 6730 \\
    bridging-data-gaps                            & 42.1 &  5.37 & 1719 &  52.9 & 35.81 &  5958 & \textbf{67.9} & 43.24 & 4865 \\
    lbcs                                          & 84.1 & 137.75 & 3369 &  46.7 & 145.55 &  6729 & \textbf{95.6} & 19.75 & 6491 \\
    sapg                                          & \textbf{35.7} & 16.09 & 2710 &  14.6 & 15.48 &  1567 & 34.8 & 28.80 & 6434 \\
    \bottomrule
  \end{tabular}}
\end{table}

\begin{table}[h]
  \caption{Per-paper scores, cost, and wall-clock time for Qwen3.6-Plus on PaperBench Code-Dev. Best score per row in bold.}
  \label{tab:qwen-paperbench}
  \centering
  \resizebox{\textwidth}{!}{
  \begin{tabular}{l rrr rrr rrr}
    \toprule
    & \multicolumn{3}{c}{\textbf{Single-Agent (100 iter)}} & \multicolumn{3}{c}{\textbf{GitWorktree (2 Eng)}} & \multicolumn{3}{c}{\textbf{STORM (2 Eng)}} \\
    \cmidrule(lr){2-4} \cmidrule(lr){5-7} \cmidrule(lr){8-10}
    Paper & Score & Cost & Time & Score & Cost & Time & Score & Cost & Time \\
    \midrule
    pinn                                               & 72.5 & 1.64 & 1860 & 77.4 & 5.99 & 4671 & \textbf{81.2} & 4.92 & 6043 \\
    lbcs                                               & 43.3 & 2.66 & 2138 & 77.2 & 17.72 & 6564 & \textbf{88.5} & 2.90 & 4631 \\
    sample-specific-masks                              & \textbf{78.7} & 2.35 & 2499 & 75.9 & 4.89 & 4847 & 68.7 & 3.20 & 4604 \\
    stochastic-interpolants                            & 74.7 & 1.63 & 1553 & 71.7 & 4.79 & 3313 & \textbf{76.8} & 2.47 & 2525 \\
    what-will-my-model-forget                          & 27.4 & 1.84 & 2785 & \textbf{69.2} & 4.17 & 5456 & 55.2 & 3.39 & 2079 \\
    mechanistic-understanding                          & 70.6 & 2.36 & 2536 & 65.0 & 6.39 & 6377 & \textbf{84.4} & 3.86 & 6889 \\
    bridging-data-gaps                                 & 30.0 & 1.88 & 2366 & \textbf{62.9} & 7.84 & 6070 & 50.2 & 5.96 & 3078 \\
    lca-on-the-line                                    & 52.1 & 2.90 & 2712 & \textbf{59.2} & 17.85 & 8140 & 42.9 & 5.08 & 3180 \\
    sequential-neural-score-estimation                 & 67.8 & 3.26 & 3345 & 56.6 & 2.82 & 8542 & \textbf{71.6} & 3.42 & 2344 \\
    stay-on-topic-with-classifier-free-guidance        & \textbf{58.8} & 2.05 & 3007 & 50.6 & 6.43 & 4746 & 52.9 & 3.13 & 3420 \\
    bam                                                & 76.7 & 2.72 & 1565 & 50.5 & 5.40 & 4819 & \textbf{76.9} & 9.87 & 6708 \\
    fre                                                & 31.2 & 1.59 & 1326 & 50.4 & 5.27 & 3624 & \textbf{53.0} & 6.73 & 3556 \\
    ftrl                                               & 25.2 & 2.31 & 2491 & \textbf{48.7} & 11.43 & 6813 & 43.7 & 7.18 & 3204 \\
    adaptive-pruning                                   & 23.4 & 3.21 & 1912 & \textbf{43.1} & 5.85 & 5563 & 31.6 & 4.28 & 3885 \\
    test-time-model-adaptation                         & 54.4 & 2.66 & 3600 & 40.6 & 4.15 & 4191 & \textbf{64.7} & 7.11 & 3368 \\
    rice                                               & 33.4 & 2.29 & 1708 & \textbf{36.7} & 5.17 & 4094 & 22.4 & 3.96 & 2147 \\
    robust-clip                                        & 13.5 & 2.03 & 1266 & \textbf{30.5} & 5.04 & 4202 & 29.7 & 4.15 & 2035 \\
    all-in-one                                         & \textbf{59.5} & 2.84 & 2614 & 28.1 & 5.33 & 6073 & 47.1 & 2.60 & 1480 \\
    bbox                                               & 21.4 & 1.96 & 1715 & 20.1 & 5.65 & 6254 & \textbf{42.4} & 2.41 & 1248 \\
    sapg                                               & \textbf{39.3} & 2.60 & 3220 & 18.3 & 5.63 & 7000 & 15.2 & 3.42 & 2967 \\
    \bottomrule
  \end{tabular}}
\end{table}

\begin{table}[h]
  \caption{Per-paper scores, cost, and wall-clock time for DeepSeek-V4-Pro on PaperBench Code-Dev. Best score per row in bold.}
  \label{tab:deepseek-paperbench}
  \centering
  \resizebox{\textwidth}{!}{
  \begin{tabular}{l rrr rrr rrr}
    \toprule
    & \multicolumn{3}{c}{\textbf{Single-Agent (100 iter)}} & \multicolumn{3}{c}{\textbf{GitWorktree (2 Eng)}} & \multicolumn{3}{c}{\textbf{STORM (2 Eng)}} \\
    \cmidrule(lr){2-4} \cmidrule(lr){5-7} \cmidrule(lr){8-10}
    Paper & Score & Cost & Time & Score & Cost & Time & Score & Cost & Time \\
    \midrule
    sample-specific-masks                              & 82.8 & 1.79 & 3600 & \textbf{90.8} & 7.22 & 8200 & 82.8 & 7.71 & 7715 \\
    pinn                                               & 87.4 & 2.07 & 3536 & 88.4 & 6.79 & 4730 & \textbf{88.9} & 4.66 & 3745 \\
    what-will-my-model-forget                          & 73.1 & 2.69 & 3600 & \textbf{82.3} & 2.65 & 5533 & 79.7 & 6.37 & 8477 \\
    lbcs                                               & 72.5 & 2.01 & 3601 & 78.0 & 3.73 & 4596 & \textbf{95.8} & 5.64 & 6477 \\
    stochastic-interpolants                            & \textbf{81.3} & 1.22 & 2513 & 69.0 & 3.57 & 4813 & 60.2 & 4.51 & 5717 \\
    sequential-neural-score-estimation                 & 82.5 & 2.48 & 1873 & 68.7 & 4.30 & 5973 & \textbf{84.2} & 5.52 & 8880 \\
    bbox                                               & 39.2 & 2.42 & 3600 & \textbf{65.6} & 4.33 & 4136 & 53.3 & 4.71 & 6289 \\
    bam                                                & \textbf{91.9} & 4.37 & 3600 & 65.3 & 5.57 & 7399 & 79.1 & 6.48 & 7407 \\
    fre                                                & \textbf{65.0} & 2.58 & 2112 & 64.4 & 8.38 & 8058 & 55.8 & 8.77 & 8355 \\
    all-in-one                                         & 87.8 & 1.87 & 3570 & 63.8 & 9.28 & 6804 & \textbf{97.6} & 8.63 & 8035 \\
    test-time-model-adaptation                         & 60.2 & 2.34 & 1875 & \textbf{62.9} & 2.86 & 4533 & 57.2 & 3.61 & 5615 \\
    adaptive-pruning                                   & 37.6 & 3.28 & 3600 & 47.1 & 6.79 & 8182 & \textbf{53.5} & 4.98 & 5644 \\
    bridging-data-gaps                                 & 45.7 & 3.54 & 3600 & 45.7 & 6.15 & 6034 & \textbf{58.1} & 5.80 & 6667 \\
    ftrl                                               & 41.5 & 2.80 & 3600 & \textbf{45.1} & 4.81 & 5377 & 43.6 & 8.77 & 6679 \\
    mechanistic-understanding                          & 76.1 & 2.42 & 3226 & 42.8 & 2.53 & 5204 & \textbf{88.3} & 8.05 & 6821 \\
    lca-on-the-line                                    & 47.1 & 2.80 & 3408 & 36.5 & 4.44 & 6418 & \textbf{75.8} & 8.53 & 6330 \\
    robust-clip                                        & 37.0 & 2.13 & 2633 & 33.6 & 5.40 & 6135 & \textbf{65.1} & 11.68 & 8752 \\
    stay-on-topic-with-classifier-free-guidance        & 62.7 & 3.31 & 2579 & 29.7 & 3.25 & 7393 & \textbf{72.2} & 4.62 & 6227 \\
    rice                                               & \textbf{51.8} & 3.30 & 2954 & 28.1 & 2.82 & 5696 & 20.3 & 6.45 & 7823 \\
    sapg                                               & \textbf{34.1} & 2.18 & 3225 & 8.6 & 3.24 & 4474 & 19.2 & 3.51 & 9851 \\
    \bottomrule
  \end{tabular}}
\end{table}

\section{Failure analysis details}
\label{app:analysis-failure-details}

\begin{figure}[h]
  \centering
  \includegraphics[width=\linewidth]{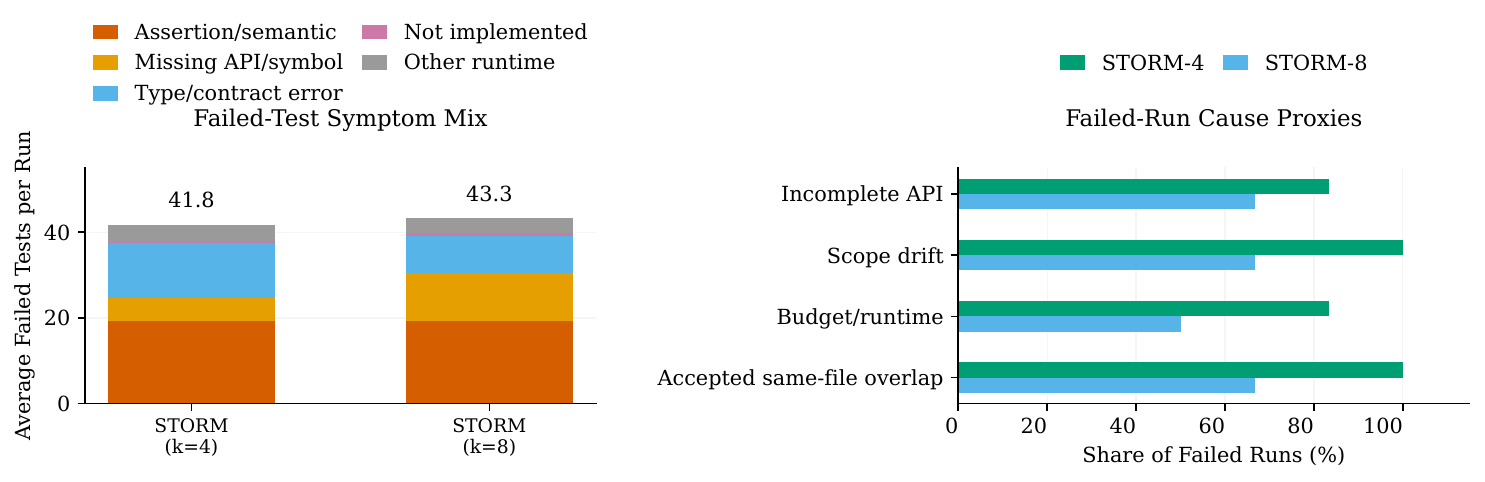}
  \caption{Failure analysis for STORM runs at $k{=}4$ and $k{=}8$. Left: average failed-test symptom mix per run, decomposed into assertion/semantic, missing API/symbol, type/contract, not-implemented, and other runtime failures. Right: share of failed runs in which each run-level cause proxy fires, covering incomplete API, scope drift, budget/runtime, and accepted same-file overlap.}
  \label{fig:failure-analysis}
\end{figure}

The failure analysis clarifies the boundary of what file-level state management can and cannot guarantee, restricted to STORM runs at $k{=}4$ and $k{=}8$. Failed tests are categorized from pytest traceback symptoms into coarse buckets: assertion or semantic failures, missing API or symbol failures, type or contract errors, not-implemented errors, and other runtime failures (Figure~\ref{fig:failure-analysis}, left). These categories are directly derived from tracebacks, but they remain symptom labels rather than manually adjudicated root causes.

Run-level cause tags are likewise heuristic proxies (Figure~\ref{fig:failure-analysis}, right). \textit{Incomplete API} is inferred from missing-symbol or not-implemented failures. \textit{Scope drift} is inferred when accepted writes modify files outside the task's declared scope. \textit{Budget/runtime} is inferred from explicit agent errors such as max-iteration or timeout failures. \textit{Accepted same-file overlap} is inferred when multiple task ids have accepted writes to the same file in a failed run. Failed tests themselves are dominated by assertion mismatches, missing APIs, and type or contract errors, indicating that unsuccessful runs still produce substantial but behaviorally incorrect implementations. Among failed STORM runs, cross-module scope drift and accepted same-file overlap appear in nearly all failed cases, and budget or runtime failures remain common at both $k{=}4$ and $k{=}8$. These proxies explain why STORM is not sufficient by itself: many remaining failures arise after writes have already been accepted as file-version consistent. STORM can ensure that accepted writes are based on a current workspace state, but it cannot guarantee that the chosen task boundaries are semantically correct, that independently accepted edits compose cleanly, or that agents complete within budget.

\begin{table}[h]
  \caption{Effect of intent annotation on Commit0-Lite with Claude Sonnet 4.6 (STORM, 4 engineers).}
  \label{tab:annotation-ablation}
  \centering
  \small
  \begin{tabular}{l cc cc}
    \toprule
    Method & Score\textsubscript{w} $\uparrow$ & Score $\uparrow$ & Cost\textsubscript{eff} $\downarrow$ & Time\textsubscript{eff} $\downarrow$ \\
    \midrule
    STORM (with annotation)    & \textbf{46.2} & \textbf{82.5} & \textbf{6.3} & \textbf{16.8} \\
    STORM (no annotation)      & 26.6 & 70.9 & 8.0 & 24.3 \\
    \bottomrule
  \end{tabular}
\end{table}

\section{Limitations}
\label{sec:limitations}

\paragraph{Terminal bypass.} STORM mediates the \texttt{file\_editor} tool but not direct filesystem writes through bash (\texttt{sed}, \texttt{echo >}, Python scripts). A post-hoc diff mechanism detects these but cannot reject them preventively.

\paragraph{No command coordination.} Concurrent shell commands (e.g., two agents running formatters on overlapping files) are not serialized. Extending concurrency control to arbitrary terminal side effects remains open.

\paragraph{File-level granularity.} Two agents editing different functions in the same file trigger a false-positive rejection. Heavily shared files (e.g., \texttt{\_\_init\_\_.py}) become serialization bottlenecks. Line-level or hunk-level tracking would reduce this at the cost of managing shifting offsets after each edit.

\section{Extended Related Work}
\paragraph{Code Generation for Software Development}
\label{sec:related-code}

LLM-based code generation has moved from function-level synthesis to broader software development tasks. Recent code generation agents can decompose tasks, write code, use tools, run tests, and debug failures~\citep{selfplanning, Rocode, ThinkLonger}. Repository-level methods further retrieve and use cross-file context from existing projects, including APIs, dependencies, tests, and coding conventions~\citep{repocoder,repofusion}. Benchmarks such as CrossCodeEval and DevEval show that realistic code generation requires cross-file reasoning and repository-level dependency understanding~\citep{crosscodeeval,deveval}.

However, code generation alone is not enough for reliable software development. Real tasks often require long-horizon interaction, repeated testing, debugging, and revision. Existing methods mainly improve planning, retrieval, or generation quality~\citep{ThinkAnywhere,li2026mem}. They provide limited support for multi-step collaboration, state control, and conflict management across agents. Execution-grounded agents add test feedback to the generation loop~\citep{agentforge, CodeScore}, but reliable coordination across long software development workflows remains an open problem.

\end{document}